\documentclass[%
reprint,
superscriptaddress,
 amsmath,amssymb,
 aps,
prb,
]{revtex4-2}

\usepackage{graphicx}
\usepackage{dcolumn}
\usepackage{bm}
\usepackage{bigints}

\begin{document}

\preprint{APS/123-QED}

\title{Application of the Green function formalism to the interplay between avalanche and multiphoton ionization induced by optical pulses}

\author{Alessandro Alberucci}
 \email{alessandro.alberucci@uni-jena.de}
\author{Chandroth P. Jisha}%
\affiliation{%
  Friedrich-Schiller-University, Institute of Applied Physics, Albert-Einstein-Str. 15, 07745, Jena, Germany}
\author{Stefan Nolte}
\affiliation{%
  Friedrich-Schiller-University, Institute of Applied Physics, Albert-Einstein-Str. 15, 07745, Jena, Germany}
\affiliation {Fraunhofer Institute for Applied Optics and Precision Engineering IOF, Albert-Einstein-Str. 7, 07745, Jena, Germany}

\date{\today}

\begin{abstract}
  A fundamental brick of light-matter interaction at large optical intensities is the generation of a plasma. The optically-induced plasma in turn plays a fundamental role in determining the optical propagation. The plasma generation is a result of the interplay between multi-photon, tunnel and avalanche ionization. Here we use the basic rate equations to discuss an analytical model for the interaction between these  physical effects. After defining a nonlinear impulse response for the system, we describe how the interplay depends on the features of the optical pulses. Our approach strongly simplifies the modelling of the propagation of ultrashort-pulses, paving the way to a much easier and faster interpretation of experimental observations, with potential impact on the broad fields of ultrafast light-matter interaction and laser micro-machining.
\end{abstract}

\maketitle


The interaction of intense optical pulses with matter is a topic of central interest in physics. Indeed, the nonlinear optical regime is intrinsically a strongly out-of-equilibrium system due to the rapidity in the exchange of energy between the electromagnetic field and the atoms \cite{Corkum:1993,Disa:2021}. This allows the experimental investigation of new regimes in condensed matter physics, including many body problems and Floquet systems \cite{Chu:2004,Oka:2019}, or the measurement of the material properties, using for example high harmonic generation \cite{Goulielmakis:2022,Lakhotia:2020,Neufeld:2023}. Beyond the basic physics, the problem is of primary importance because strong lasers can modify the properties of a material in a temporary manner \cite{Lakhotia:2020,Disa:2021} or by inducing permanent modifications, a phenomenon widely exploited in laser micro-machining \cite{Gattass:2008,Tunnermann:2023}. 

One common feature of the interaction between intense light and matter is the formation of plasma \cite{Raizer:1966,Shen:1984}. The impinging photons provide energy to the electrons of the material, thus inducing a considerable amount of electronic transitions towards higher energy states. This process takes place even in materials which are transparent in the linear regime due to the tunnel ionization (TI) \cite{Geissler:1999} and multi-photon ionization (MPI) \cite{Nathan:1985}. Once free carriers are generated, the optical field is accelerating them, providing on average an increase in the kinetic energy. The accelerated electrons can then collide with other less energetic electrons, inducing a field-dependent and concentration-dependent amplification. Such effect is called avalanche ionization (AI), and it is often associated with the dielectric breakdown \cite{Sparks:1981}.

On a theoretical ground, the problem of the strong-coupling between light and matter can be solved quantum-mechanically using the TDDFT (Time Dependent Density Functional Theory) \cite{Otobe:2020}, yet a very demanding approach from the computational point of view. At a larger scale, the dynamics of the sea of electrons subject to an optical field can be solved using the machinery of the Boltzmann's transport equations \cite{Kroll:1972}. In most of the cases, this method is prohibitive given it requires the full knowledge of the energetic dispersion for the electrons and of the loss mechanisms (e.g., excitons and electron-phonon coupling). 
A common and prolific approach is to define a distribution for the excited electrons $n_e(\bm r,t)$ averaged over the energetic and momentum states, thus depending only on space and time. This approximation works well in a wide range of materials, including liquids \cite{Kennedy:1995}, amorphous solids \cite{Sudrie:2002}, and semiconductors \cite{Fedorov:2016}.
Neglecting electron diffusion in space, $n_e$ is then dictated by the rate equation \cite{Nathan:1985,Stuart:1995,Li:1999}
\begin{equation} \label{eq:electron_distribution_MPA}
    \frac{\partial n_e}{\partial t}=  W_\mathrm{PI}(I)  +   \left(\alpha_\mathrm{av} I - \frac{1}{\tau_{el}} \right) n_e - \sigma n_e^2.
\end{equation}
In Eq.~\eqref{eq:electron_distribution_MPA} $I(\bm r,t)$ is the optical intensity of a field with central frequency $\omega$. The first term on the RHS (Right Hand Side) $W_\mathrm{PI}$ is the photo-ionization rate (PI) as predicted by the Keldysh's theory for atomic transitions under the influence of a periodic field \cite{Keldysh:1965,Nathan:1985}; the second term on the RHS $\alpha_\mathrm{av}I n_e$ accounts for the electrons excited by the avalanche effect \cite{Sparks:1981,Li:1999};  the third term $-n_e/\tau_{el}$ accounts for the average lifetime of the excited electrons due to the various recombination mechanisms. Finally, the last term proportional to the square of the density represents the nonlinear (with respect to the electron density $n_e$) recombination effects, such as Auger. 

The Keldysh's theory is amazingly capable of modelling both tunnelling and multi-photon ionization: the transition between the two regimes is demarcated by the so-called Keldysh parameter $\gamma\propto \frac{\omega}{q} \sqrt{ m c n \epsilon_0 E_g/I}$, where $E_g$ is the bandgap of the material, $c$ is the speed of light, $\epsilon_0$ is the vacuum dielectric permittivity, and finally $m$ and $q$ the mass and the charge of the electron \cite{Schaffer:2001}. In the MPI case $\gamma$ is large, in turn yielding $W_\mathrm{PI}\approx\frac{1}{\hbar \omega}\sum_{N}{ \frac{\alpha_N}{N} I^{N} }$, where $\alpha_N$ is the cross-section for the ionization involving $N$ photons. As shown below, the type of ionization (TI or MPI) does not significantly change our results, in fact providing only a different mathematical relationship between the optical intensity and the source of electrons. Thus, for simplicity, we will restrict ourselves at first on MPI and neglect TI. At the end of the Article we will show how our approach works when the full Keldysh formula is applied. Incidentally, we are for now considering the general case of multiple multi-photon transitions, although usually the smallest one fulfilling $N\hbar \omega \geq E_g$ is the relevant one.  

The usage of a rate equation to model the plasma formation in strong optical fields was already described by Shen in his seminal book about nonlinear optics \cite{Shen:1975}, and later expanded to its final form by Kennedy in 1995 \cite{Kennedy:1995}. More complicated versions introducing different energetic states have been discussed in literature \cite{Rethfeld:2006,Tsibidis:2023}, but the fundamental physics does not strongly depend on that. Furthermore, as discussed for example in Ref.~\cite{Feng:1997}, the optical breakdown is usually not strongly affected by $\sigma n_e^2$, which can then be neglected.  Under the assumptions made, Eq.~\eqref{eq:electron_distribution_MPA} is linear with respect to the electronic distribution $n_e$: it can then be solved using the Green's function formalism
\begin{equation} \label{eq:ne_integral}
    n_e(\bm r, t) = \frac{1}{\hbar \omega} \sum_{N}\frac{\alpha_N}{N} \int_{-\infty}^ \infty{  
 I^N(\bm r,t^\prime) G(\bm r, t,t^\prime) dt^\prime},
\end{equation}
where the impulse response is
\begin{equation} \label{eq:green}
    G(\bm r, t,t^\prime) = e^{\alpha_\mathrm{av}\int_{t^\prime}^t I(\bm r,\tau) d\tau } e^{-(t-t^\prime)/\tau_{el}} u_0(t-t^\prime).
\end{equation}
In Eq.~\eqref{eq:green} $u_0$ is the Heaviside function, necessary to fulfill the causality condition. The Green function also retains the reciprocity property given that $G(\bm r,t,t^\prime)=G(\bm r,t^\prime,t)$. Equations~(\ref{eq:ne_integral}-\ref{eq:green}) are the core of the reasoning and findings we are developing in this paper. In this form and as anticipated earlier, it is clear that the form of the field-induced ionization solely changes the forcing term in Eq.~\eqref{eq:electron_distribution_MPA}, thus not affecting the solution method we are proposing here. As a matter of fact, solutions of Eq.~\eqref{eq:electron_distribution_MPA} in terms of integral were already sketched in the original paper by Kennedy \cite{Kennedy:1995}, and explicitly written by Feng and collaborators in Ref.~\cite{Feng:1997}. Nonetheless, such integral solution has not been explored in depth to discuss the interplay between MPI and AI; indeed, verbatim from Ref.~\cite{Feng:1997}: \textit{``Since the analytic solution (8) is not very informative, we have numerically solved the density equation''}. Oppositely to the reported view, we show that writing such an integral in terms of the Green formalism permits to disclose how field ionization (either MPI or TI) and AI are working together and explore the underlying physical behavior. 

The first advantage of our approach is the possibility to clearly distinguish the origin of the excited electrons, and how MPI and AI interact with each other. To further simplify the notation, hereafter we will focus on the case when only one single multi-photon transition is relevant: the generalization to multiple simultaneous transitions is straightforward. From Eqs.~(\ref{eq:ne_integral}-\ref{eq:green}) the net generation of excited electrons per unit time is
\begin{align} \label{eq:generation_ne}
    \frac{\partial n_e}{\partial t} &= \frac{\alpha_N}{N\hbar \omega} \biggl[  I^N(t) \ + 
  \left(\alpha_\mathrm{av} I(t) - \frac{1}{\tau_{el}} \right) \int_{-\infty}^t I^N(t^\prime) \times
 \nonumber \\ &   e^{\alpha_\mathrm{av}\int_{t^\prime}^t I(\tau) d\tau } e^{-(t-t^\prime)/\tau_{el}}  dt^\prime  \biggl].
\end{align}
Equation~\eqref{eq:generation_ne} allows immediate physical interpretation: the excitation rate is the sum of the instantaneous MPI (first term on the RHS) plus  the avalanche electrons generated at each instant normalized with respect to the lifetime $\tau_{el}$ (the integral term). The seed for the avalanche electrons is provided by the MPI at former times, whereas the history of the pulse intensity $I(t)$ determines the overall amplification for each electron generated by MPI. Although Eq.~\eqref{eq:generation_ne} is somehow trivial under our model based upon the temporal Green function, it can not be easily extracted from numerical solutions of Eq.~\eqref{eq:electron_distribution_MPA}.

We now turn our attention to describe the type of response modelled through Eq.~\eqref{eq:green}. For the sake of simplicity, hereon we will omit the spatial dependence which is not relevant in our current discussion. Once a shape for the pulse is fixed, the Green function depends only on the product $\alpha_\mathrm{av}I_0$, where $I_0$ is the intensity peak. The shape of the response of the material depends on the relative position along the pulse profile [i.e., $G(t,t^\prime)\neq G(t-t^\prime)$] through the avalanche term, i.e., the response is not invariant with respect to time shifts; accordingly, the distribution $n_e$ is not given by a simple temporal convolution. On a more physical ground, Eq.~\eqref{eq:ne_integral} is telling us that $n_e$ at a given instant $t$ is the sum of the electrons excited by MPI at each previous instants, but such electron density needs to be weighted with respect to the amount of amplification -fixed by the net balance between avalanche and losses- the electrons have been subject to. From Eq.~\eqref{eq:green}, the position $t_\mathrm{max}$ of the extrema of the Green function $G$ versus $t$ is 
\begin{equation} \label{eq:max_Green}
   I(t_\mathrm{max}) = \frac{1}{\tau_{el}\alpha_\mathrm{av}}.
\end{equation}
The causality condition imposes the additional constraint $t_\mathrm{max}>0$.
The former equation holds valid irrespective of the temporal shape $I(t)$. Generally speaking, Eq.~\eqref{eq:max_Green} correctly predicts that, for larger $\alpha_\mathrm{av}$ or for longer electron lifetime $\tau_{el}$, the maximum of the avalanche-generated electrons shifts towards the trailing edge of the pulse, regardless of when the seed electrons (i.e., $t^\prime$) have been generated.


The fluence is the temporal integral of the intensity $I$. Defining the time-windowed fluence $F(t_1,t_2)$ as the amount of fluence between the two instants $t_1$ and $t_2$, Eq.~\eqref{eq:green} provides
\begin{equation} \label{eq:green_series}
    G(t,t^\prime) = \left[ 1 + \alpha_\mathrm{av}^n \sum_{n=1}^\infty{ \frac{F^n\left(t^\prime,t \right)}{n!} } \right] e^{-(t-t^\prime)/\tau_{el}} u_0(t-t^\prime).
\end{equation}
The optical response (free electrons actually change both the imaginary and the real part of the refractive index) of the material can then be controlled by shaping the impinging pulses, with potential applications in the novel field of photonic materials encompassing a time-dependent response \cite{Galiffi:2022,Tirole:2023}. Furthermore, as the intensity is ramping up, more terms in Eq.~\eqref{eq:green_series} become relevant: the joint action of MPI and AI effectively behaves like a multi-photon ionization of order $N+n$ \cite{Rajeev:2009}, but encompasses an additional memory effect registering the previous slices of the optical pulses already passed through the material \cite{Conti:2010}. \\
We now focus on how the MPI and AI interact. Given that Eq.~\eqref{eq:green_series} provides the amount of electrons excited by an impulsive pulse placed in $t=t^\prime$, the avalanche process excites more electrons than the MPI once
\begin{equation} \label{eq:transition}
     \lim_{t\rightarrow\infty}F(t^\prime, t) > \frac{1}{\alpha_\mathrm{av}}.
\end{equation}
This means that the share of AI-excited electrons with respect to the ones ascribed to MPI depends primarily on the fluence which crossed the material after the excitation instant $t^\prime$. Hence, the shape of the pulse -including the pulse duration $\tau$- determines the (continuous) transition between the two regimes; the transition between MPI and AI occurs at a given instant $t^\prime$ along the optical pulse. When the maximum fluence $F(-\infty,\infty)$ is lower than $1/\alpha_\mathrm{av}$, MPI remains the dominant mechanism exciting the electrons to the conduction band. \\
\begin{figure}[!ht]
\centering\includegraphics[width=0.49\textwidth]{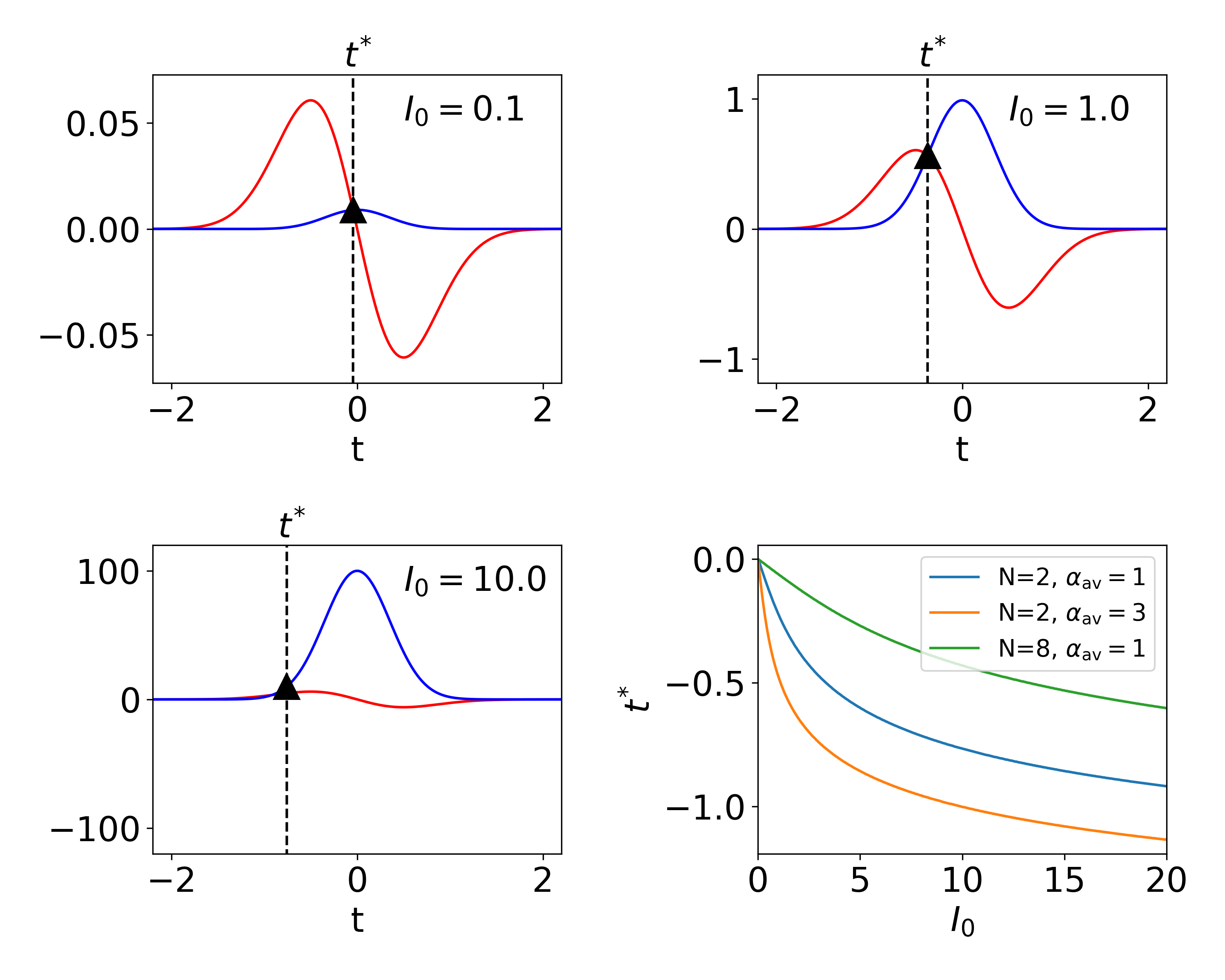}
\caption{(a-c) Graphical solution of Eq.~\eqref{eq:max_condition} for a Gaussian pulse featuring a pulse duration $\tau=1$ and a peak intensity $I_0$ equal to 0.1 (a), 1 (b), and 10 (c). Blue and red curves correspond to the LHS and RHS of Eq.~\eqref{eq:max_condition}, respectively. In (a-c) we assumed $\alpha_\mathrm{av}=2$. (d) Behavior of $t^*$ versus $I_0$ for three different pairs of $\alpha_\mathrm{av}$ and $N$ as labelled in the legend. Here we have fixed $\tau_\mathrm{el}=50$.} 
\label{fig:sketch_tmax}
\end{figure}

Actually, Eq.~\eqref{eq:transition} alone does not ensure the dominance of AI over MPI in the ionization process. Indeed, AI could be dominant at the front edge of the pulse, where the low intensity generates  a modest number of seed electrons via the MPI to be later accelerated by the avalanche process. To assess this matter, we can calculate which instant $t^\prime$ (we dub it $t^*$) contributes for the largest number of electronic transitions. By deriving the integrand of Eq.~\eqref{eq:ne_integral} and setting such time-derivative equal to zero, we find the condition
\begin{equation} \label{eq:max_condition}
    \frac{\partial I}{\partial t} = \frac{\alpha_\mathrm{av}}{N} I^2 - \frac{I}{N\tau_{el}}.
\end{equation}
For simplification, let us assume a single-humped pulse shape, see Fig.~\ref{fig:sketch_tmax} for a graphical solution of the equation \cite{Hunter:2007,Harris2020}. The RHS of Eq.~\eqref{eq:max_condition} needs to be positive to ensure $t^*<0$ to achieve net gain, thus setting the constraint $I_0>1/\left(\alpha_\mathrm{av}\tau_{el}\right)$ with $I_0$ being the maximum intensity. In fact, in the absence of avalanche ($\alpha_\mathrm{av}=0$) and for $\tau_{el}\rightarrow\infty$, the maximum of $n_e$ corresponds to the  intensity peak.
When $\alpha_\mathrm{av}$ is large enough, the instant $t^*$ moves towards the front edge of the pulse, where the temporal derivative $\partial I/\partial t$ [RHS of Eq.~\eqref{eq:max_condition}] is non-vanishing and positive, see the red curves in Fig.~\ref{fig:sketch_tmax}(a-c). Thus, larger intensities enhances the shift of $t^*$ towards earlier instants, the larger the $\alpha_\mathrm{av}$ the larger the shift is [compare the blue and orange curves in Fig.~\ref{fig:sketch_tmax}(d)].  Finally, greater $N$ favors the MPI by decreasing the temporal shift of $t^*$ with respect to lower $N$ [compare the blue and green curves in Fig.~\ref{fig:sketch_tmax}(d)].

We now proceed with showing applications of Eqs.~(\ref{eq:ne_integral}-\ref{eq:green}) for specific optical pulse shapes: the scope is to demonstrate the versatility of our approach in determining the influence of the pulse shape on the plasma generation. We make the additional assumption that the shape of the optical pulse is fixed: we are thus neglecting self-phase modulation, both in space and time \cite{Schaffer:2001}.
To be more quantitative and provide closed-form solutions for $n_e$, we now suppose the pulse to be a square function, $I(t)=I_0 \text{rect}_\tau(t)$, where the $\text{rect}$ function is non-vanishing and equal to 1 only for $|t|<\tau/2$. After defining the net gain $g\left(I_0\right)=\alpha_\mathrm{av}I_0-1/\tau_{el}$, Eq.~\eqref{eq:ne_integral} with the help of Eq.~\eqref{eq:green_series} provides 
\begin{align} \label{eq:ne_rect_pulse}
    n_e(t) &= \frac{\alpha_N I_0^N}{g(I_0) N\hbar \omega} \biggl\{ \left[e^{g(I_0)\left(t+\frac{\tau}{2}\right)}-1\right]u_0\left(-t+\frac{\tau}{2}\right)   \biggr\} \nonumber \\ & + n_e^{max} e^{-\frac{t}{\tau_{el}}} u_0\left(t-\frac{\tau}{2}\right) .
\end{align}
In Eq.~\eqref{eq:ne_rect_pulse} we also defined the peak of the electron density as
\begin{equation} \label{eq:ne_max_rect_pulse}
    n_e^{max} = \frac{\alpha_N I_0^N}{g(I_0) N\hbar \omega} \left[e^{g(I_0)\tau} -1 \right].
\end{equation}
The interpretation of Eq.~\eqref{eq:ne_rect_pulse} is straightforward: during the pulse the number of electrons grows exponentially with the intensity-dependent net gain $g(I_0)$. The density $n_e$ achieves its maximum $n_e^{max}$ at the end of the pulse ($t=\tau/2$), then exponentially decays with a lifetime determined by $\tau_{el}$. The interplay between AI and MPI can be evaluated by expanding the exponential term in Eq.~\eqref{eq:ne_max_rect_pulse} in its power series. At low gains ($g(I_0)\tau\ll 1$), we obtain $n_e^{max}= \alpha_N I_0^N \tau /(N\hbar\omega)$, i.e., the case of electrons generated only by MPI. In the opposite limit $g(I_0)\tau\gg 1$, we have a purely exponential growth of the first electrons generated at the leading edge of the pulse, $n_e^{max}\propto \left. W_{PI}\right|_{t=-\tau/2} e^{g(I_0)\tau}/g(I_0)$. For the intermediate case we expand the exponential series up to the quadratic terms, providing the following condition for the transition to an avalanche-dominated ionization
\begin{equation} \label{eq:transition_rect_pulses}
    \alpha_\mathrm{av}I_0 > \frac{2}{\tau} + \frac{1}{\tau_{el}},
\end{equation}
in agreement with Eq.~\eqref{eq:max_condition}. \\
Thus, in the case of square pulses the transition between MPI and AI does not depend on the transition order $N$: the shortest between the pulse duration $\tau$ and the electron lifetime $\tau_{el}$ is actually determining the transition between the two regimes. In agreement with the numerical simulations of Eq.~\eqref{eq:electron_distribution_MPA}, AI becomes dominant after a given threshold intensity dependent on the pulse duration, with shorter pulses favouring MPI. Finally, from Eq.~\eqref{eq:ne_max_rect_pulse} it is expected that the nonlinear absorption increases as $I_0^N$ for small enough intensities, then depending as $I_0^{N+1}$ when AI kicks in, finally going to a full exponential increase when AI is largely dominant. This general trend is upper bounded in real experiments by the optical breakdown and permanent modifications induced in the material. 

\begin{figure}[!ht]
\centering\includegraphics[width=0.49\textwidth]{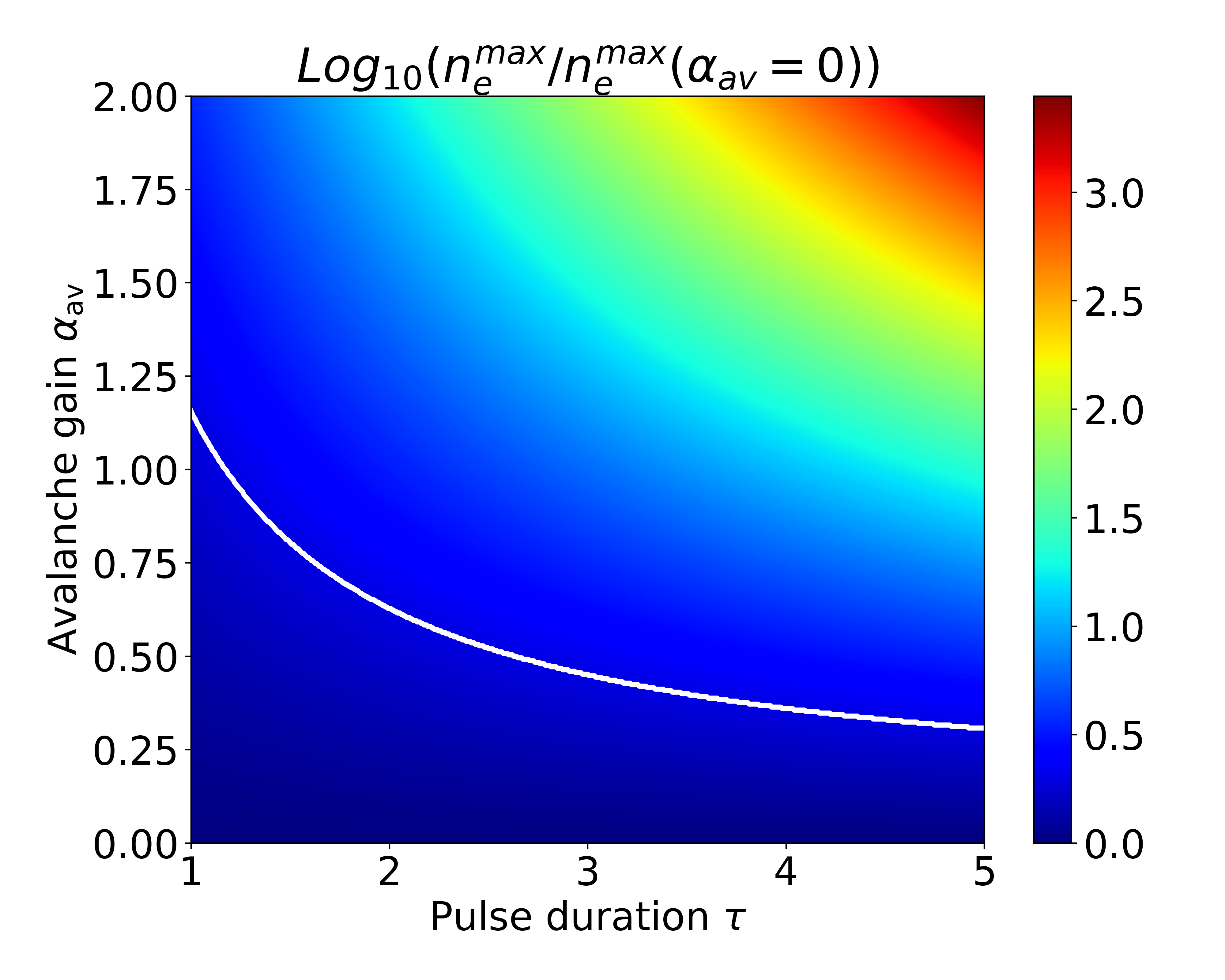}
\caption{Maximum of the electron density versus the AI coefficient $\alpha_\mathrm{av}$ and the pulse duration $\tau$ for Gaussian pulses featuring $I_0=1$. The density are normalized with respect to the density without avalanche and plotted in logarithmic scale. The white solid line is the condition when the $n_e^{max}$ is doubled with respect to the case $\alpha_\mathrm{av}=0$. Here we fixed $I_0=1$, $N=2$ and $\tau_{el}=50$.} 
\label{fig:overall_gain}
\end{figure}
\begin{figure}[!ht]
\centering\includegraphics[width=0.49\textwidth]{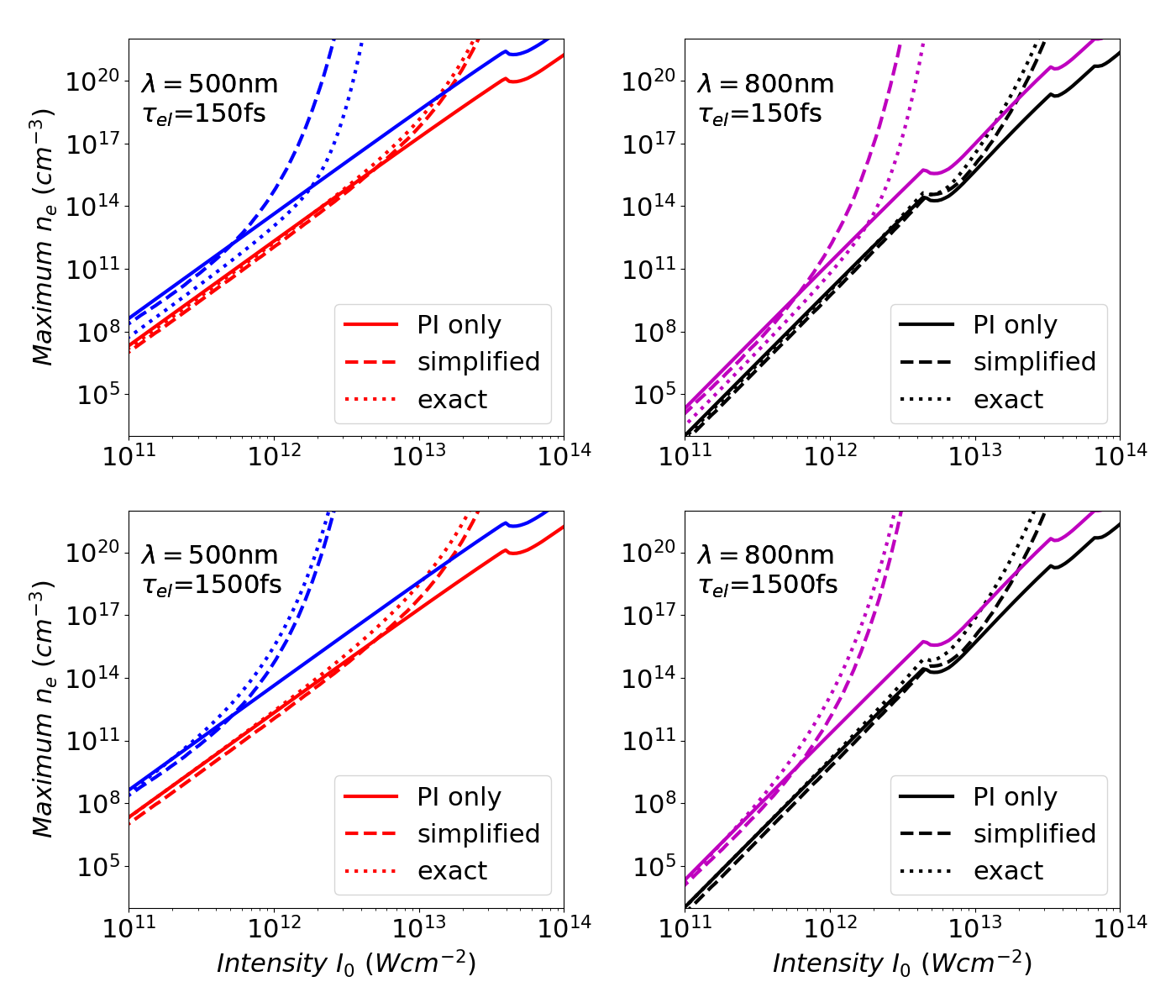}
\caption{Maximum of the electron density $n_e$ versus the peak intensity $I_0$ in fused silica. Wavelength is $500~$nm (left column) and $800~$nm (right column). Electron lifetime is $\tau_{el}$ is $150~$fs (top row) and $1.5~$ps (bottom row). Red and black (blue and magenta)  curves correspond to a pulse duration $\tau$ of $150~$fs ($2~$ps).} 
\label{fig:ne_fused_silica}
\end{figure}

We now pass to the most common case of a Gaussian-shaped pulse, $I=I_0e^{-2t^2/\tau^2}$. The function $F$ is then $F(t,t^\prime)=\frac{I_0\tau }{2} \sqrt{{\frac{\pi}{2}}} \left[\text{erf}\left( \frac{\sqrt{2}t}{\tau}\right) - \text{erf}\left( \frac{\sqrt{2}t^\prime}{\tau} \right) \right]$. From Eq.~\eqref{eq:max_Green} we get 
\begin{equation} \label{eq:max_Green_gaussian_pulse}
    t_\mathrm{max} = \tau \sqrt{\frac{\log\left(\tau_{el}\alpha_\mathrm{av}I_0\right)}{2}}.
\end{equation}
\begin{figure}[!ht]
\centering\includegraphics[width=0.49\textwidth]{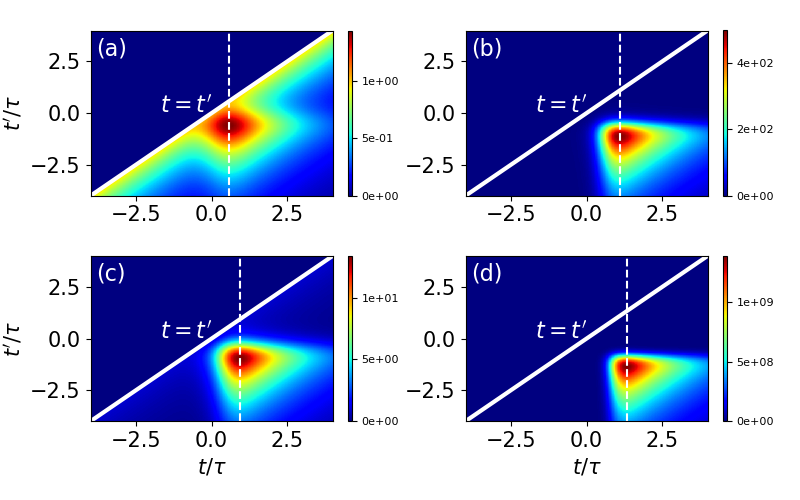}
\caption{Green function for a Gaussian pulse centered in $t=0$. The pulse duration $\tau$ is 1 (a,b) and 6 (c,d). The avalanche amplification is 1 (a,c) and 6 (b,d), whereas we fixed $I_0=1$ and  $\tau_{el}=2\tau$ in all the panels. Vertical dashed lines is the maximum position according to Eq.~\eqref{eq:max_Green_gaussian_pulse}.} 
\label{fig:green_gaussian_pulse}
\end{figure}
Figure~\ref{fig:green_gaussian_pulse} shows the Green function for two pulse durations and two values of $I_0\alpha_\mathrm{av}$, where we fixed $\tau_{el}=2\tau$. The peak of $G$ versus $t$ is always positioned after the peak of the pulse, i.e., $t_\mathrm{max}>0$. Due to the exponential amplification, the peak of $G$ steeply grows for larger products $I_0\alpha_\mathrm{av}$ (comparison between columns) and  for longer pulses  (comparison between different rows), as well known both from numerical simulations and experiments \cite{Kennedy:1995}. When $\tau_{el}$ is much longer than the pulse duration $\tau$ (e.g., femtosecond pulses), the Green function does not drop significantly for increasing time. On the opposite limit $\tau_{el}\ll \tau$ (e.g., nanosecond pulses), the electrons accumulation is hindered, with the the maximum of $G$ migrating towards earlier instants, see Eq.~\eqref{eq:max_Green_gaussian_pulse}. \\
\begin{figure}[!ht]
\centering\includegraphics[width=0.49\textwidth]{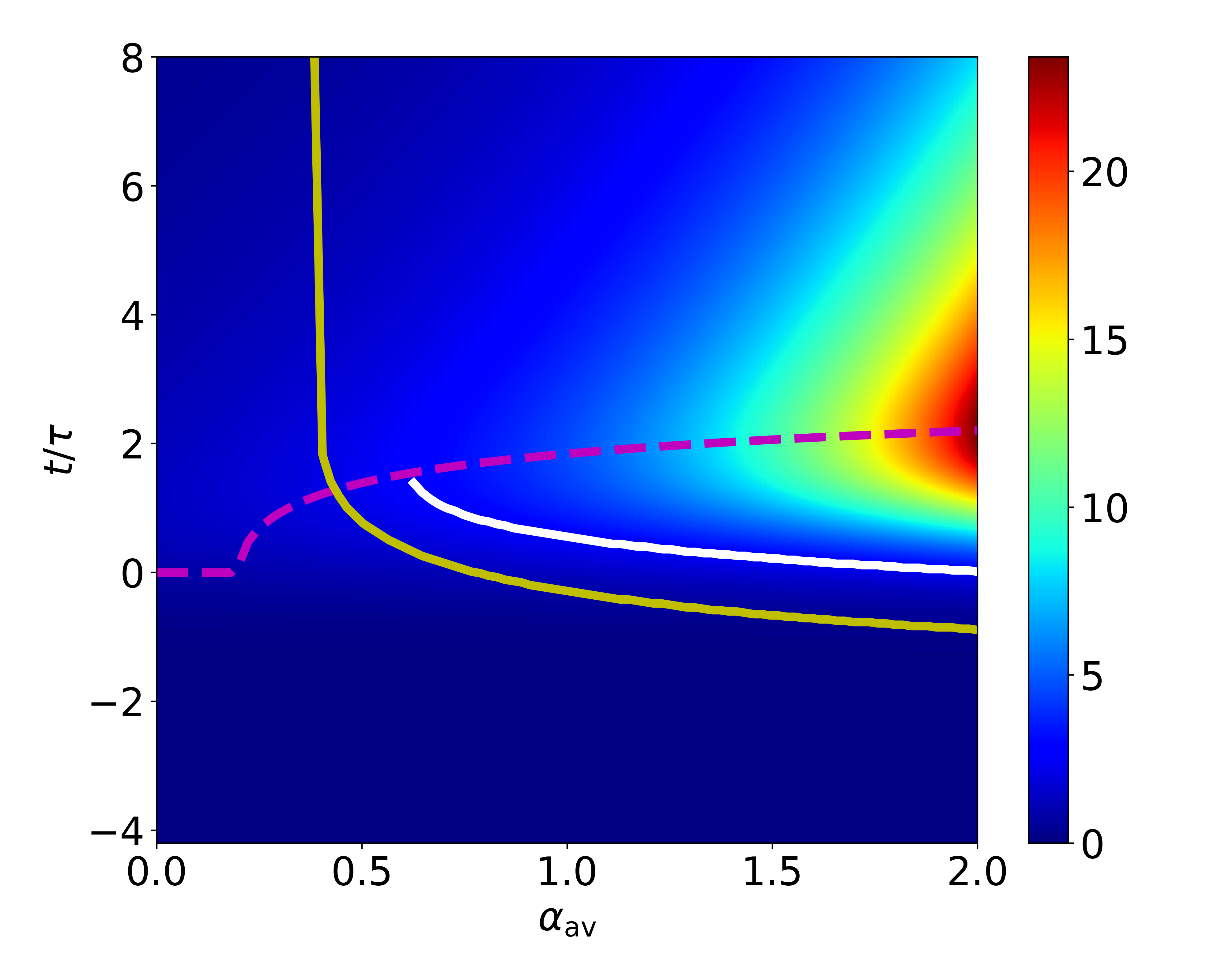}
\caption{Electron density $n_e$ versus the normalized time $t/\tau$ (vertical axis), parameterized with respect to the avalanche coefficient $\alpha_\mathrm{av}$ (horizontal axis). The magenta dashed line is the position of the peak given by Eq.~\eqref{eq:max_Green_gaussian_pulse}. The white solid line is where the electron density becomes double the excitation due to MPI alone. The yellow solid line is the set of instants satisfying Eq.~\eqref{eq:transition}. The used missing parameters are $I_0=1$, $\tau=2$, $N=2$ and $\tau_{el}=50$.} 
\label{fig:ne_vs_t}
\end{figure}
The results of the integration of Eq.~\eqref{eq:ne_integral} for Gaussian pulses and $\alpha_N=1$ is shown in Fig.~\ref{fig:ne_vs_t}. In agreement with the shape of the Green function, $n_e$ first reaches a maximum after the pulse peak, and then drops with a rate determined by $\tau_{el}$. The maxima of $n_e$ (magenta dashed line) are always placed at $t=t_\mathrm{max}$, the latter corresponding to the peak of the Green functions, regardless of $t^\prime$. With respect to $\alpha_\mathrm{av}$, the shape of $n_e$ versus $t$ does not substantially changes, except for an exponential amplification. The interplay between MPI and AI can be first evaluated by finding the temporal instants where the density doubles with respect to the maximum of $n_e$ calculated when $\alpha_\mathrm{av}=0$ (white solid line in Fig.~\ref{fig:ne_vs_t}). Such condition is achieved only when $\alpha_\mathrm{av}$ overcomes a given threshold, strongly dependent on the other parameters of the pulse. After achieving the threshold, the curve follows a hyperbola-like trend. For the sake of comparison with the theory developed above, we draw in the same graph the points where the condition defined by Eq.~\eqref{eq:transition} is satisfied (yellow solid line). The two curves are almost parallel, with the theoretical prediction being slightly more stringent (i.e., AI dominance at earlier times and lower avalanche coefficients) than the numerical one. \\  
Next, we investigate how the maximum electron density $n_e^{max}$ varies for different pulse durations $\tau$ and avalanche coefficients $\alpha_\mathrm{av}$. Typical results are shown in Fig.~\ref{fig:overall_gain}. In qualitative agreement with Eq.~\eqref{eq:ne_max_rect_pulse}, the electron density grows exponential both with $\tau$ and $\alpha_\mathrm{av}$. The transition to an avalanche-driven process -defined as a doubling of $n_e$ as in Fig.~\ref{fig:ne_vs_t}- is represented by the white solid line in Fig.~\ref{fig:overall_gain}. In full analogy with Eq.~\eqref{eq:transition_rect_pulses}, the curve is hyperbolic, but with a coefficient now dependent on the multiphoton order $N$.\\ 
We now apply our model to a real case, fused silica illuminated by optical beams emitting at two different wavelengths, $\lambda=500$~nm and $\lambda=800~$nm. Whereas up to this point our calculations were carried out in normalized units, we now pass to physical units.
From Ref.~\cite{Couairon:2005} we take the parameters $\alpha_\mathrm{av}=4\times 10^{-4}$m$^2$J$^{-1}$ and $E_g=9~$eV, in turn providing $N=4$ at $\lambda=500~$nm and $N=6$ at $\lambda=800~$nm. 
For the photo-ionization we employ the full Keldysh formula, thus accounting for the transition from MPI to TI as the impinging intensity increases \cite{Couairon:2005}. 
Figure~\ref{fig:ne_fused_silica} shows the corresponding maximum in the electron density versus the peak intensity $I_0$ for Gaussian pulses of different widths and different wavelengths. The solid lines represent the predicted value in the absence of avalanche and $\tau_{el}\rightarrow \infty$. The use of the full Keldysh formula is responsible for the abrupt changes in the distribution. In agreement with the multiphoton ionization, the slope is steeper for longer wavelength due to the larger $N$. When avalanche is accounted for, a sudden exponential growth in $n_e$ is taking place. Before such a divergence, the two cases match well for long enough electron lifetime $\tau_{el}$. Furthermore, the generation of electrons is stronger for longer pulses for a fixed peak intensity $I_0$. In particular, the avalanche amplification is strongly enhanced for longer pulse durations, in agreement with the literature and Eq.~\eqref{eq:ne_max_rect_pulse}. The dashed lines in Fig.~\ref{fig:ne_fused_silica} (labelled as \textit{simplified} in each panel) is the amount of electrons generated starting from the seed induced by PI only at $t=t^*$, see Eq.~\eqref{eq:max_condition} and Fig.~\ref{fig:sketch_tmax}. When $\tau\ll \tau_{el}$, the exact number of excited electrons is a little bit larger, but the trend with $I_0$ is almost identical. When $\tau_{el}$ is shorter than the pulse duration $\tau$, the approximation is overestimating the real amount of excited electrons. In agreement with Eq.~\eqref{eq:transition_rect_pulses}, the onset of avalanche for a given peak intensity is determined by the interplay between electronic lifetime and pulse duration. In the case of Gaussian pulses, solution of Eq.~\eqref{eq:transition_rect_pulses} is a very good proxy for determining the transition to an avalanche-dominated regime.\\
As a last result, we aim to apply our approach to discuss how the simultaneous illumination of the material with two pulses of different duration affects the plasma generation. 
Recently, the relevant role of temporal contrast in determining the modifications in bulk materials has been investigated experimentally, both in fused silica and in silicon \cite{Wang:2020,Wang:2023}.  We consider two pulses illuminating the sample simultaneously: a short [dubbed $I_s(t)$] and a long pulse [dubbed $I_l(t)$] with a duration of 150~fs and 10~ps, respectively. We also assume that the two pulses are perfectly synchronized by placing the peak of both pulses at $t=0$. In applying the Keldysh formula, we simply used the sum of the two intensities: thus, we are neglecting possible coherent interference during the transitions between the excited electronic states, see e.g. the coherent control \cite{Silberberg:2009}. Eq.~\eqref{eq:green} provides
\begin{equation} \label{eq:green_joint}
    G_\mathrm{joint}(t,t^\prime) = e^{\alpha_\mathrm{av}\int_{t^\prime}^t \left[ I_s(\bm r,\tau) + I_l(\bm r,\tau)  \right] d\tau } e^{-(t-t^\prime)/\tau_{el}} u_0(t-t^\prime),
\end{equation}
that is, the total avalanche gain is simply given by the multiplication of the separate gains, at least until Eq.~\eqref{eq:electron_distribution_MPA} holds valid.
The electron density is
\begin{equation} \label{eq:ne_joint}
    n_e(t) =  \int_{-\infty}^ \infty{  
 \ W_\mathrm{PI}(I_s +I_l) G_\mathrm{joint}(t,t^\prime) dt^\prime}.
\end{equation}
\begin{figure}[!ht]
\centering\includegraphics[width=0.49\textwidth]{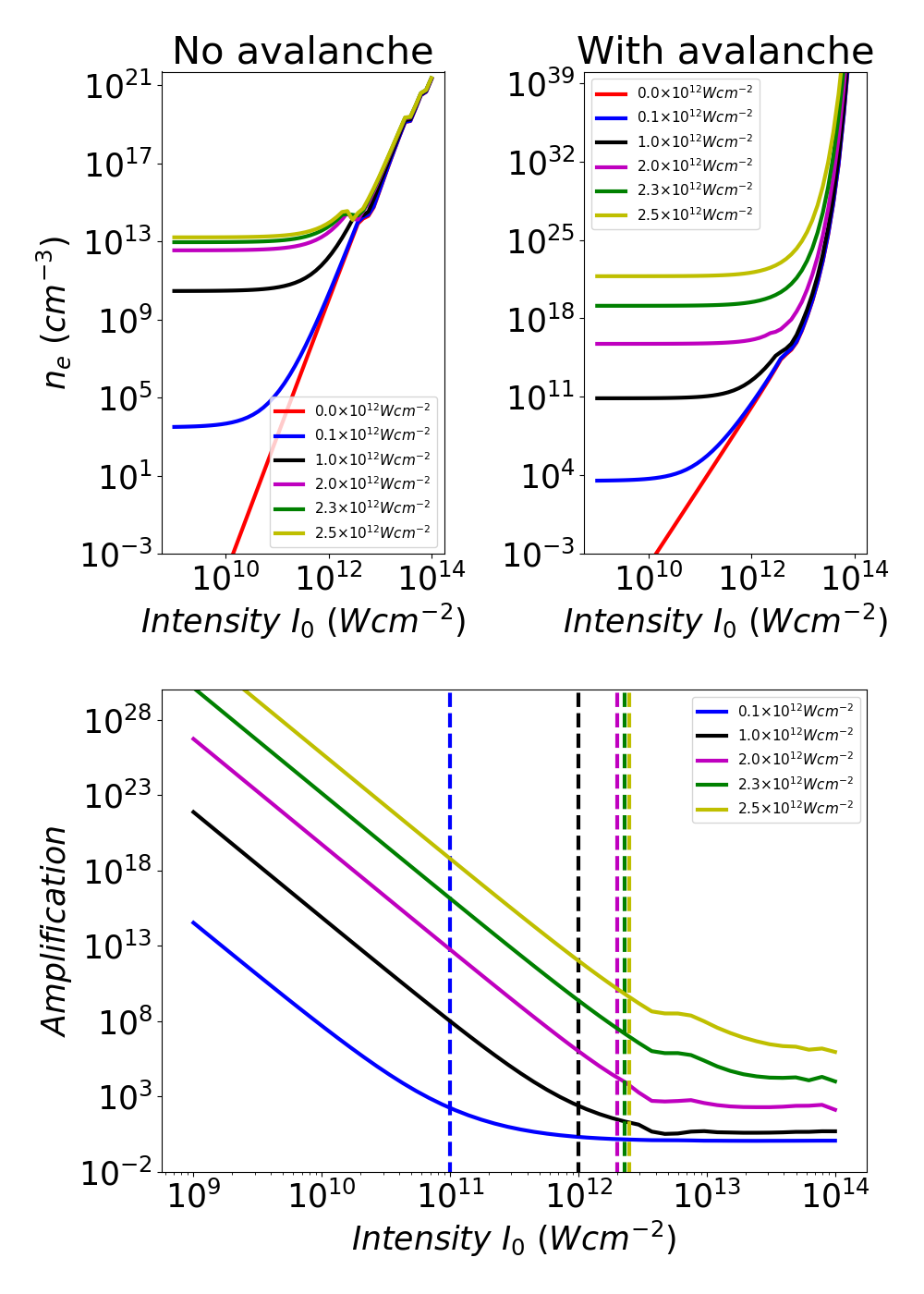}
\caption{Top row: Maximum of the electron density $n_e$ in fused silica versus the peak intensity $I_0$ of the short Gaussian pulse with duration 150~fs without (left side) and with (right side) avalanche. Bottom row: ratio between the electron densities with and without the long pulse and in the presence of avalanche. The vertical dashed lines show when the two pulses have the same peak intensity. Each color corresponds to a different peak intensity (see legend) of the long pulse with duration 10~ps. Wavelength is 800~nm and $\tau_{el}=150~$fs.} 
\label{fig:joint_pulses}
\end{figure}
Figure~\ref{fig:joint_pulses} compares the maximum electron density achieved for different values of the peak intensities, without (top left panel) or with (top right panel) avalanche (note the different scaling on the vertical axis). The horizontal axis is the peak intensity of the short pulse $I_s(t)$, whereas each curve corresponds to a different peak intensity for the long pulse $I_l(t)$. Important to stress, here the peak intensity, which actually depends on the ratio between the pulse energy and the pulse duration, is kept fixed.
We start discussing the case without avalanche, which provides the electrons excited by direct field ionization (either MPI or TI). 
The joint curve differs from the case of isolated pulses only when the intensities are comparable, in agreement with the Keldysh formula.
When avalanche is turned on, the exponential gain caused by the short pulse alone does not significantly change  in the presence of a long pulse with an intensity up to $2.0\times 10^{-12}$~Wcm$^{-2}$ and lower than the short pulse  (see the blue and the black line in the bottom row of Fig.~\ref{fig:joint_pulses}).  When $I_l$ approaches the threshold for avalanche, the exponential gain strongly differs from the gains calculated when only $I_s$ is illuminating the material: indeed, the amplification, defined as $n^{max}_e/n_e^{max}(I_l=0)$ and plotted in the bottom of Fig.~\ref{fig:joint_pulses}, does not saturate to unity when $I_s$ gets larger and larger. Essentially, using two pulses it is possible to decouple the plasma density from the intensity amplitude and shape, the latter being non-separable in the case of single-pulse illumination.  If the long pulse is anticipated with respect to the short pulse with a delay small with respect to $\tau_{el}$, such a scheme can be used to study the interaction between a tunable density of electrons (fixed by the long pulse, assumed to not induce permanent modifications in the material) and a short pulse of variable intensity. 



In conclusion, we introduced an analytical model based upon the Green function formalism to depict the excitation of high-energy electrons in the presence of nonlinear photo-ionization (i.e., multiphoton and tunnel) and avalanche ionization. The model allows a versatile and rapid investigation on how many electrons are excited for a given optical pulse, in fact providing a clear picture of the interplay between different ionizations and its dependence on the parameters of the optical pulse. 
Due to its simplicity, the model can be readily integrated in more advanced algorithms computing the optical propagation in the nonlinear regime, such as beam propagation method codes or approximated solutions based upon the variational theory. Experimentally, our model is a fast and efficient tool to describe pump-probe set-ups measuring the temporal dynamics of the absorption after illumination with an intense pulse \cite{Bergner:2018,Jurgens:2019}. With respect to applications, our method represents a rapid solution for estimating the best parameters to inscribe permanent structures in solids. Finally, our approach paves the way to the employment of pulse shaping to control the electron density  for inputs in proximity of the onset of avalanche ionization \cite{Wang:2020}.

\begin{acknowledgments}

Supported by the Free State of Thuringia and the European Social Fund Plus (2022FGR0002).
European Union’s Framework Programme for Research and Innovation Horizon 2020 under the Marie Sklowdowska-Curie Grant Agreement No. 889525.
\end{acknowledgments}




\providecommand{\noopsort}[1]{}\providecommand{\singleletter}[1]{#1}%

\end{document}